\documentclass{article}
\usepackage{latexsym}
\usepackage{amsmath,amssymb}
\usepackage{float}
\usepackage[dvipdfmx]{graphicx}%

\begin{document}

\pagestyle{plain} 
\setcounter{page}{1}
\setlength{\textheight}{650pt}
\setlength{\topmargin}{-40pt}
\setlength{\headheight}{0pt}
\setlength{\marginparwidth}{-10pt}
\setlength{\textwidth}{20cm}

\title{Cooperation in Evolutionary Public Goods Game on Complex Networks with Topology Change}
\author{Norihito Toyota and Takumi Takahashi \and Faculty of Business Administration and Information Science, \\ \hspace{6mm} Hokkaido Information University, Ebetsu, Nisinopporo 59-2, Japan\\
toyota@do-johodai.ac.jp }
\date{}
\maketitle
%
%



\abstract{
 The evolution of cooperation among unrelated individuals in human and animal societies remains a challenging issue across disciplines. 
It is an important  subject also in the evolutionary game theory to research how cooperation arises. 
The subject has been extensively studied  especially in Prisonars' dilemma game(PD) and 
the emergence of cooperation is important subject also in public goods game(PGG).  
 In this article, we consider evolutionary PGG on complex networks where the topology of the networks varies under the infulence of game dynamics.  
Then we study what effects on the evolution of player's strategies, defection and cooretation and the average payoff does the interaction between the game dynamics and the network topology bring. 
By investigating them by making computer simulations, we intend to clear in what situations cooperation strategy is promoted or preserved. 
We also intend to investigate the influence of the interaction on the average payoff over all players. 
Furthermore how  initial  networks are transformed to final networks by the evolution through the influences of PGG dynamics is investigated. 
\subsubsection*{keyword; public goods game, complex networks, evolutionary game, \\ \hspace{17mm}cooporation,  network evolution }


\section{Introduction}
The evolution of cooperation among unrelated individuals in human and animal societies remains a challenging issue across disciplines. 
Since the 1990s, it has been an important  subject also in the evolutionary game theory\cite{Smi} to research how cooperation arises. 
The subject has been extensively studied during the past thirty years, especially in Prisonars' dilemma game(PD) and its various modified versions \cite{Now}.
A rational player chooses Nash equilibrium strategy, defection, in the original version of the one-off PD with two strategies by two playsers\cite{Wei}.  
So by what mechanism cooperation, that is Pareto Optimum, is realized, it is a question.  
In the iterated games of PD, a tendency to choose the Nash equilibrium strategy is relieved under some conditions\cite{Ax}. 
There are many circumstantial evidences that the cooperation is promoted within groups in the evolutionary PD on lattice, which is a sort of spcial games\cite{Now},\cite{Now2}.
Similar studies for evolutionary games on complex networks\cite{New}, which are considered as metaphors for various relations in the real world, in addition to regular networks such as lattice, have been investigated\cite{Sza},\cite{Dor},\cite{Santos1}. 
Moreover it is begun to pay its attention to the relations between the evolution of cooporation and the network structure in social dilemma games\cite{Perc1}\cite{Holme}. 
The recearchs on how does game dynamics on networks influence the network topology, has been developed\cite{Fu},\cite{Gross}. 
The PD on growing networks\cite{Pon1}\cite{Pon2}, time-varying networks and other networks \cite{Car}  for promoting cooperation have been studied.  
 Recently it is pointed that the evolution of strategies may not  satisfyingly propmte cooperative behavior in  the evolutionary games. 
It has been recognized that the effect of  spatial structure and heterogeneity in addition to the evolution of strategies is important for cooperation \cite{Perc2},\cite{Perc3},\cite{Perc4}. 
In the coevolution in the evolutionary games another property besides the evolution of strategies may simultaneously be subject to  evolutions in general.  
See an excellent review by \cite{Perc5} for researches about such coevolution. 
Furthermore it has been pointed out that the combination of network structure and other factors such as memory promotes cooperation behavior\cite{Hadz1}, \cite{Hadz2}. 
However, its assertion are not yet conclusive and sufficiently general, because the analysis of interactions between network topology and game dynamics is rather intricated\cite{Toyo}. 

On the other hand, the Public Goods Game(PGG) \cite{Now},\cite{Mc},\cite{Hau} that can be interpreted as $n$-persons PDhas been considered.  
Studies for cooperation as have been made in PD game have already begun  also in PGG\cite{Bat}.  
Especially PGG on complex networks with topology change has been studyed\cite{Wu1},\cite{Wu2},\cite{Zhang1},\cite{Zhang2}, \cite{Sha},\cite{Perc6}. 
First PGG using the normalized enhancement factor $\delta$  on random networks\cite{Ren} has been studied\cite{Wu1}.  
A focal player compares his/her payoff  with the payoff of a player chosen randomly among neighbors  and change the strategy based on Fermi rule where either a player experience strategy dynamics with a probability $w$ or network dynamics is undergone with a probability $1-w$. 
In \cite{Wu2}, PGG on Barabasi-Albert network\cite{Bara} has been studied. 
There the network is fixed but a focal player plays PGG with  $g$ players, which is a fixed number,  among neighbors of a focal player.   
In the model, comparing the focal player's payoff  with the payoff of a player chosen randomly among neighbors,  his/har strategy is also changed  and, the change of the network structure is based on "reputation" of players\cite{Wu2}. 
PGG on lattice has been studied in \cite{Zhang1}. 
In their model, players break the adverse social ties bringing the worst productivity, whereas remain ties with others. 
After breaking the ties, the player will search for a more beneficial interaction with another partner among the neighbors of neighbors.
There  a measure of individual's inertia to react to their rational selections both at strategy and topological levels is introduced. 
The strategy and the local network structure of a site are updated with probabilities $\omega$ and $1-\omega$, respectively. 
Different $\omega$ values separate those so that the two time scales is implemented by asynchronous updating $\omega$. 
The model in \cite{Zhang2} has explored PGG on Newman-Watts small world network\cite{New2}. 
A charcteristic property of their model is that  the change of network topology is undergone based on "aspiration level",  whereas the strategy is undergone based on Fermi rule. 
The model proposed by \cite{Sha} introduce "aspiration" for strategy change and the change of network topology where an unsatisfied  player with negative satisfaction defined by using the aspiration breaks an edge to randome neighbor, changes strategy or add an edge to a random player.   
The model is only based on the player's personal information but doesn't need to know any information of the neighbors of 
the player.   
All of the works assert that cooperation can be promoted or maintained within the framework of their models. 
A fine review including rich references is described in \cite{Perc7}.     

  In this article, we consider evolutionary PGG on diverse complex networks where the topology of the networks varies under the infulence of game dynamics.  
There the strategy change and the change of network topology are so simple in the sense that our models are stchastic ones but do not introduce any outside factors such as the aspiration (level), Fermi rule and  reputation. 
We introduce some strategy models that are considered to reflect and simplify people's features  in the real world. 
The models sometimes include a part of futures of above-mentioned models\cite{Wu1},\cite{Wu2},\cite{Zhang1},\cite{Zhang2}, \cite{Sha},\cite{Perc6}.  
We will study in a lump including a part of them in this article. 
 The some models  reflect game dynamics whereas others  do not. 
Though diverse complex networks are investigated as initial networks, we mainly study how does the interaction between the game dynamics and the network topology influence the evolution of player's strategies, defection and cooretation, and the average payoff over all players.   
Investigating them by making computer simulations, we intend to clarify in what situations cooperation is promoted or preserved. 
We also intend to investigate the infuluence of the interaction on the average payoff over all players. 
Furthermore we investigate how  initial  networks are transformed to final networks by the network evolution through the influences of PGG dynamics.      

The structure of this article is as follows. 
In the section 2, we explain the original PGG and the PGG on a network. 
After that we introduce PGG on complex network as our models.  
There we introduce evolutionary models with diverse tactics including network evolution. 
The some models  reflect game dynamics whereas others  do not. 
Then the comparison of  both type models would uncover the influences of game dynamics on coopperation or affluence. 
Even more some models change the network topology every game round where the network evoluves, whereas others do not. 
The comparison of  both type models would uncover the influences of topology change of the network on coopperation or affluence. 
We present the results of simulations  of the poroposed models in the section 3.
After performing extensive computer simulations, we select and present only the most conclusive results in this article. 
In the last section, we briefly summarize our main findings.   

 \section{PGG on Network and Models}
 \subsection{PGG}
PGG is a game with two strategies that puts into a public pot (C-strategy, cooperators) or not(D-strategy, defectors) for  public benefit.
 Players secretly choose how many of their private tokens to put into the public pot. 
The tokens in this pot are multiplied by a factor $r$ (synergy factor), where $r$ is normally greater than one and less than the number of players $n$, and this public good payoff is evenly divided among all players.  So the tokens are kept even in players who do not contribute. 
The group's total payoff is maximized when everyone contributes all of their tokens to the public pool. 
The Nash equilibrium in this game, however, is not to contribute by all. 
Those who contribute below average or nothing are called "free riders"\cite{Hau}. 

In this article, we assume that tokens to put into a public pot are a constant $M>0$. 
So cooperators gain the following payoff $P_c$;
\begin{equation}
P_c=\frac{rmM}{n}-M=M(\frac{rm}{n}-1)
\end{equation}
 where $m$ is the number of cooperators. 
 The payoff $P_d$ gained by defectors who cotribute nothing is
 \begin{equation}
P_d=M\frac{rm}{n}.
\end{equation}
 So rational players always choose D-strategy because $P_c< P_d$. 
However, all players gain more profit at large $m$ than small $m$ in general.  
The larger the number of cooperators becomes, the larger payoffs gained by players become.   
That is a dilemma. 
This dillemma is analogous to Prisoners' dilemma where the Paret optimal is not Nash equilibrium. 
For simplicity, we set $M=1$ in this article. 

\subsection{PGG on Networks}
We extend the original PGG to PGG on networks in this subsection\cite{Santos2}. 
For example, let's consider a network given in Fig.1, where players exist on nodes. 
Nodes connected with a node  by edges represent  opponents who the player on the node plays PGG. 
Players mutually connected by edges mean that there are some personal relations  among them.   
In Fig.1, player A plays PGG with players B, C, D and E. 
This PGG is called A-chair game, which is played by 5 players.  
On the other hand, player D plays PGG with 2 players, A and E. 
So in Fig.1 player A  have to play 5 games, that are  B, C, D and E-chair games in addition to A-chair game.  
The total payoff  of player A is the sum of the 5 games. 
In general, every player i plays  $k_i+1$ PGGs, where $k_i$ is the degree of node i. 

\subsection{Strategy Models}
We propose some models for strategy evolution.
They can be broadly classified into the following two types. 
One is that a player playing many PGGs gererally employs different strategies for each game, which are represented by  a strategy vector and called A-Model.  
Another  is that a player playing many PGGs employs the same strategy for each game, which is called B-Model. \\

A-Model: Every player i plays $k_i+1$ games with different strategies. 
So every player has a strategy vector $S^A_i$;
\begin{equation}
S_i^A=( s_{i1}^A, s_{i2}^A, s_{i3}^A, \cdots ,s_{i n}^A), 
\end{equation}
where $s_{ij}^A=\{0,1\}$.
 $s_{i j}^A=1$ when the player does put a token into a public pot and  $s_{ij}^A=0$ when the player 
doesn't. 
The dimension of the vector $S_i^A$ is fixed to the number of players, $n$, so that 
 we keep the dimension of the strategy vectors  for all players at constant value. 
We have to prepare the strategies even for unconnected players because  the opponents for a player
may change moment by moment due to topology change of the netorwork. 
The strategy vectors for all players are integrated into a matrix; $n \times n$ Strategy Matrix; $\bf S^A$.    
 \\

B-Model:Every player i plays $k_i+1$ games with the same strategies, C or D. 
So a player chooses a fixed strategy but strategies for all players are integrated into a vector $S^B$;
\begin{equation}
S^B=(s_1^B, s_2^B, s_3^B,\cdots ,s_n^B), 
\end{equation}
where $s_i^B=\{0,1\}$.
 $s_{i}^B=1$ when the player does put a token into a public pot and  $s_{i}^B=0$ when the player 
doesn't.   

\begin{figure}[t]
\centering
	\includegraphics[width = 6.0cm]{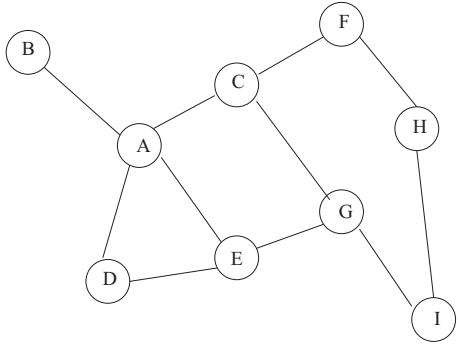}
	\caption{PGB on Network .}
\end{figure}

 Initially we take $s_i^A=1$ or $s_i^B=1$ with the probability $c$ that is a model parameter and $s_i^A=0$ or $s_i^B=0$ with the probability $1-c$. 
Some updateing models for strategies are obtained for A-model and B-model, respectively. 
We introduce two models for tactics to explore the effects of the game dynamics for A-Model. \\

A-1 Model is the model that a player i imitates the strategy vector of the player who accepted the largest payoff
amonng players connected with the player i in the next round. 
Thus the strategies of players are affected by the game dynamics. 
They are greedy persons.  \\

A-2 Model is the model that a player i increases the ratio of C(D) by $t$ in his/her strategy vector, 
when the total number of C(D) strategy
in i-$th$ component of the strategy vectors of the players connected with the player i is bigger than the one of D(C), 
 in the next round. 
Thus the strategies of players are not affected by the game dynamics. 
They are opportunism. \\

Similarly to A-Model, three models for tactics are introduced in B-Model. \\

B-1 Model is the model that a player i imitates the strategy of the player 
that accepted the largest payoff amonng players connected with the player i in the next round. 
Thus the strategies of players are affected by the game dynamics like A-1 Model. 
They are also greedy persons.  \\

B-2 Model is the model that a player i chooses the strategy of the majority side among 
the  players connected with the player i, in the next round. 
Thus the strategies of players are not affected by the game dynamics like A-2 Model. 
They are also opportunism. \\

B-3 Model is the model that a player flips his/her strategy from C(D) to D(C) in the next round, when the payoff acquired by the player becomes negative. 
They are  persons who hate to admit defeat. \\

We do not consider any model corresponding to B-3 Model in A-Model, because we think that a person 
 would not drastically change his(her) strategy 
such as flipping all the components of a strategy vector in a practical manner. 
Such thought has already seen in A-2 Model where the change of the ratio of C or D strategy in a strategy vector is conntrolled by a not so large parameter $t$.

\subsection{Topology Change  Models}
  We propose two models for topological evolutions of a network. \\
  
 $\alpha$-$Model$\\
 A player i cuts the edge with one of players with D strategy, who are harmful to the player i or reduce the payoff of  the player i,  with the probability $p$.  
 The player i sets up  a new edge to an unconnected player j chosen randomely with the probability $q$.
 
$\beta$-$Model$	\\
A player i cuts an edge with one of players with D  strategy, which are harmful to the player i or reduce the payoff of  the player i,  with the probability $p$.  
 The player i sets up  a new edge to an unconnected player j who are not directly connected but connected in two hops with the player i  with the probability $q$. \\
 
We make simulations of these models on the following typical three networks as initial networks to explore 
the transition of the average payoff and the number of cooperators, and topological transition, especially the degree distributions for all models. 

1.Random Networks(ER, Erdos-Reny model\cite{Ren}), 

2. Scale Free networks (SF, Barabashi-Albata model\cite{Bara}) with scaling exponent=3, 

3. Small World networks(WS, Wattz-Strogatz models\cite{ws} with rewriting ratio $w=0.1 \sim 0.01$), 
 where $w$=0.01 $\sim$ 0.1 is the parameter area that shows obvious small world properties. 

We should study the cases with $p=q=0$ and $p=q \neq 0 $ to explore the influence of topology change on the  average payoffs and the number of cooperators. 
At $p=q$, the total number of edges is kept nearly constant. 
The different two cases reveal the effects of topology change.

\section{Simulation Results}
In simulations, we fix $M=1$, $t=0.1$ , and take $N=200$. 
$k$ is taken between $4$ and $16$. 
$k=4 \sim 6$ is representatives of low degrees and $k=16$ is a representative of high degrees
 for $N=200$. 
 $r$ is taken between $0.5$ and $8.0$. 
 Though $r=0.5$ seems to be trivial since it is expected that all D-strategy is realize, simulations do not always show such results.  
 In this section, varying $p$, $q$, $r$, $k$ and $c$, we investigate the ratio of D-strategy, the average payoff and the future of the constructed network topology after topology change, especially degree distributions.  
$c$ has only an influence on the final converged value of D-ratio but does not have any influences on the 
whole behaviors of the time change of the D-ratio.
When  $0<\{p, q\}<1$, the results are the same as those of $p=q=1 $, but only the convergence  becomes slow. 
The number of repitition is appropriately chosen in the same parameters by observing
 how the results converge.

In view of such properties, we show mainly the chracteristic results of $\alpha$-model in this article. 

\subsection{A models}

\textbf{A-1model:} Fig.2 and Fig.3 show the ratio of D-strategy with and without topology change  in A-1 model in WS network with $w=0.1$ and $k=4$.
This essential behaviors are same as those of the other initial networks. 
C-strategy is promoted still more in the case with network evolution, especially large $r$. 
The D-strategy is particularly promoted at $r=1$, which is a boundary synagy whether total  token is amplified or not. 
When network topology does not change, D-strategy is almost dominant. 
In concert with these observations, we see that the average payoff is larger in the case with network evolution  at small degree in Fig.4 and Fig.5. 
When $r$ is large, the average payoff becomes large through all models. 

\textbf{A-2model:} Fig.6 and Fig.7 clearly show that the ratio of C-strategy rather increases in the case with network  evolution.  
Really  C-strategy occupys 50 percent or more of the whole for all $r$. 
Then the average payoff is also larger in low degrees as shown in Fig.8 and Fig.9.  
This fact is reflected in the fact that public goods become larger when free riders are ralely found.

\subsection{B models}
\textbf{B-1model:} All players choose the D-strategy at networks with a high degree, except at large $r$, 
regardless of with and without topology change as seen in Fig10 and Fig. 11.
 It is natural that the ratio of C-strategy at large $r$ increases more than that  at small $r$, relatively,  
since the relative difference of the payoffs between players with D-strategy and ones with C-strategy narrows at large $r$. 
Then C-strategy proves more advantageous than D-strategy. 
It is found from Fig.10 and Fig.11 that such tendency is more remarkable in the case with topology change.  

From Fig.12 and Fig.13, we observe that the strategy of all players becomes C-strategy or  D-strategy in most cases at small degree, but dependently on $r$ to some extent. 
Though it is thought that D-strategy usesually conquers C-strategy, the following phenomena can occur so that C-strategy makes a conquest of the network. 
There are first cases that some small size groups with all C-strategy that get comparatively high payoff accidentally occur and are steadied. 
When the some players of the group conected with a few players with D-strategy, there is the possibility that the palayer with C-strategy is the most well-off, especially at large $r$.  
Then C-strategy  is able to make a conquest of D-strategy.  
Secondly in the case where a network is covered almost by players with D-strategy, the payoffs of the players are almost zero. 
When there is one player with C-strategy, payoffs of the player and players connected the player are not zero and positive at large $r$. 
When there are two such groups and a player with C-strategy in a first group connects to another player with C-strategy in the second group, the two players with C-strategy may get larger payoffs than players with D-strategy in the groups. 
So the number of players with C-strategy increase, and after all players on the network may become full of C-strategy.      
Since the players with D-strategy are often cut,  players with C-strategy have relatively large degree. 
Thus the players with C-strategy trend to gain profit and so easy to spread at large $r$. 
Such situation  reinforces the above-mentioned two phenomena.   
The average payoff, of course, becomes larger when C-strategy is numerically superior to D-strategy, which is observed also from Fig.14 and Fig. 15 where the average payoff necessarily becomes zero when all players choose D-strategy. 
Such the situations are almost the same, regardless of with and without topology change. 
\begin{figure}[tbp]
 \begin{minipage}{0.5\hsize}
  \begin{center}
\includegraphics[width = 7.0cm]{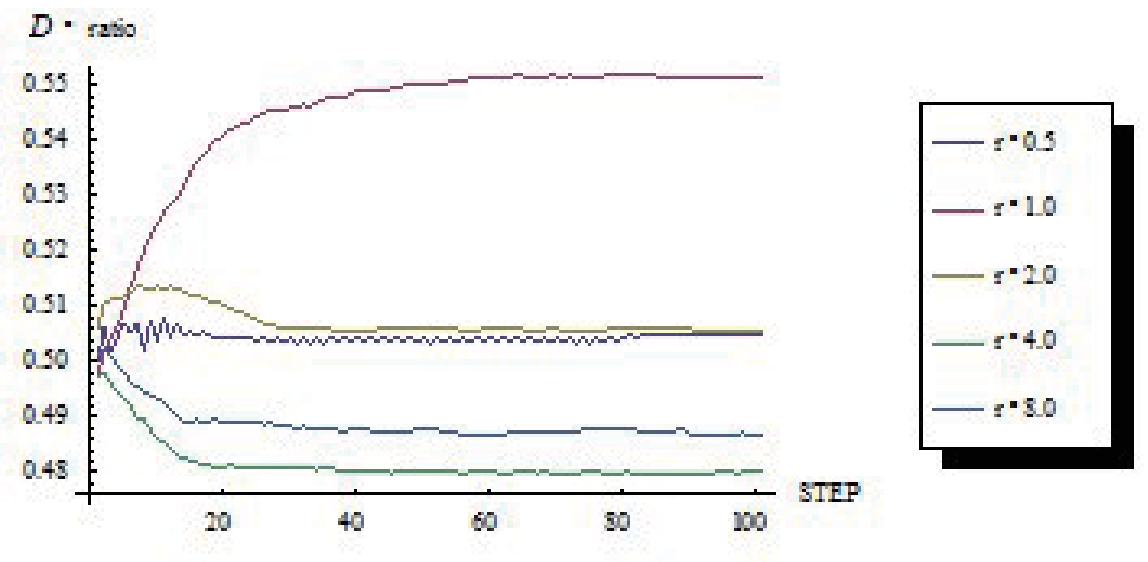}
  \end{center}
  \caption{\small The D ratio in A1-model (WSnet 
 $w=0.1$ and $k=4$) with topology change}
  \label{fig:one}
 \end{minipage}
 \hspace*{3mm}
 \begin{minipage}{0.5\hsize}
  \begin{center}
\includegraphics[width = 7.0cm]{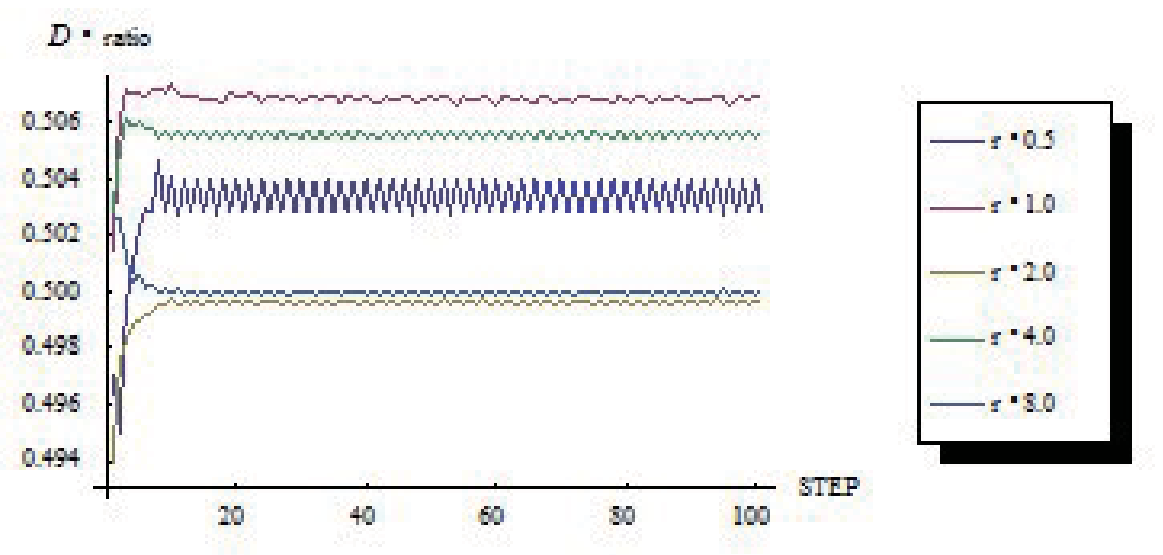}
  \end{center}
  \caption{\small The D ratio in A1-model (WSnet 
 $w=0.1$ and $k=4$) without topology change}
  \label{fig:two}
 \end{minipage}
\end{figure}

 \begin{figure}[tbp]
 \begin{minipage}{0.5\hsize}
  \begin{center}
\includegraphics[width = 7.0cm]{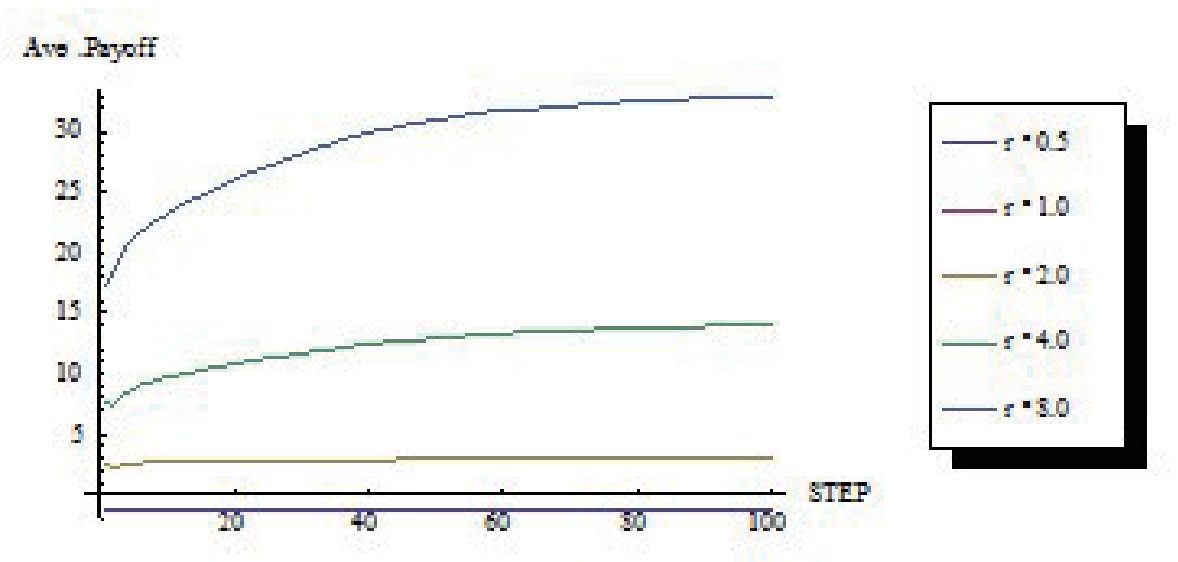}
  \end{center}
 \caption{\small The average payoff in A1-model (WSnet $w=0.1$ and $k=4$) with topology change}
  \label{fig:one}
 \end{minipage}
  \hspace*{3mm}
 \begin{minipage}{0.5\hsize}
  \begin{center}
\includegraphics[width = 7.0cm]{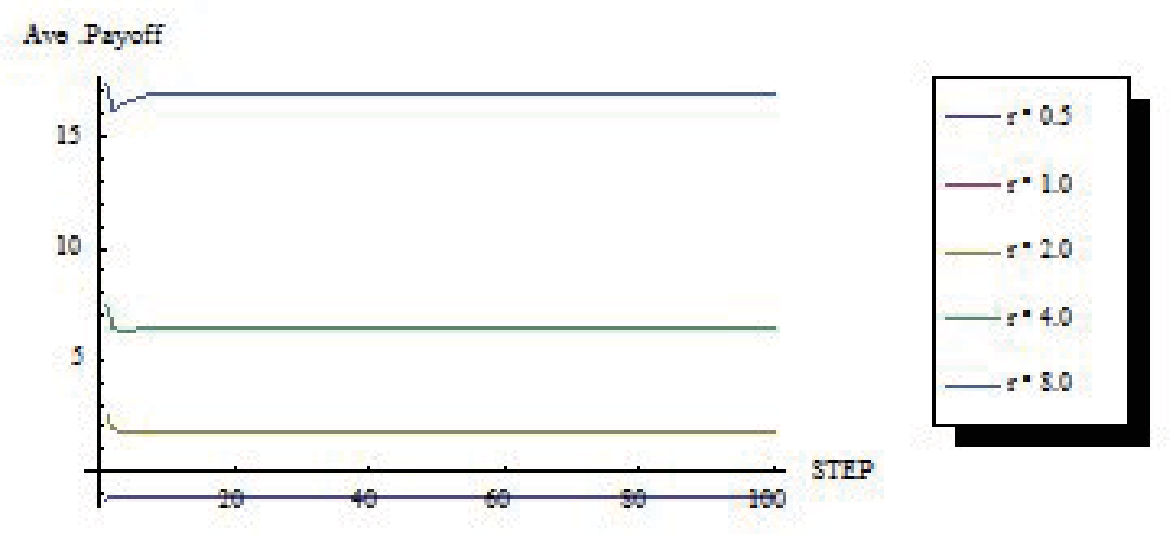}
  \end{center}
 \caption{\small The average payoff in A1-model (WSnet $w=0.1$ and $k=4$) without topology change}
  \label{fig:two}
 \end{minipage}
\end{figure}

 \begin{figure}[ttbp]
 \begin{minipage}{0.5\hsize}
  \begin{center}
\includegraphics[width = 7.0cm]{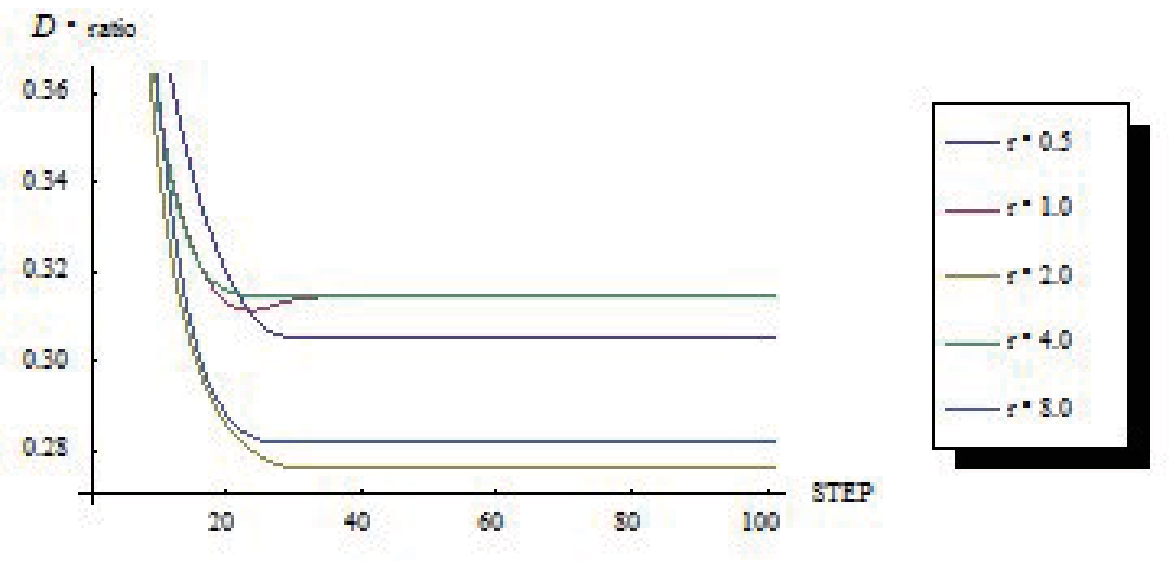}
  \end{center}
  \caption{\small The D ratio in A2-model (ERnet $k=4$) with topology change}
  \label{fig:one}
 \end{minipage}
  \hspace*{3mm}
 \begin{minipage}{0.5\hsize}
  \begin{center}
\includegraphics[width = 7.0cm]{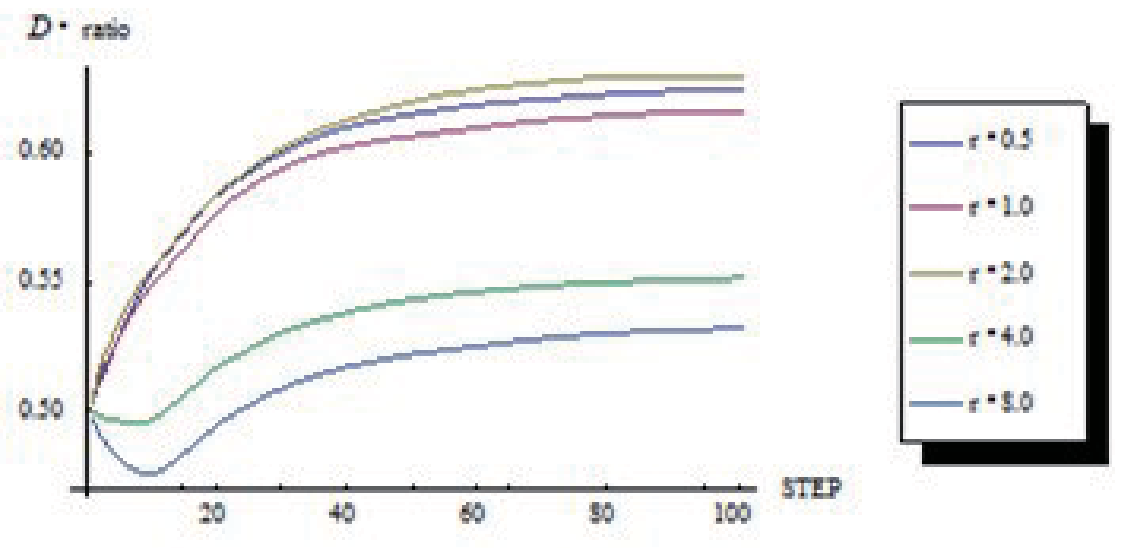}
  \end{center}
   \caption{\small The D ratio in A2-model (ERnet $k=4$) without topology change}
  \label{fig:two}
 \end{minipage}
\end{figure}

 \begin{figure}[tbp]
 \begin{minipage}{0.5\hsize}
  \begin{center}
\includegraphics[width = 7.0cm]{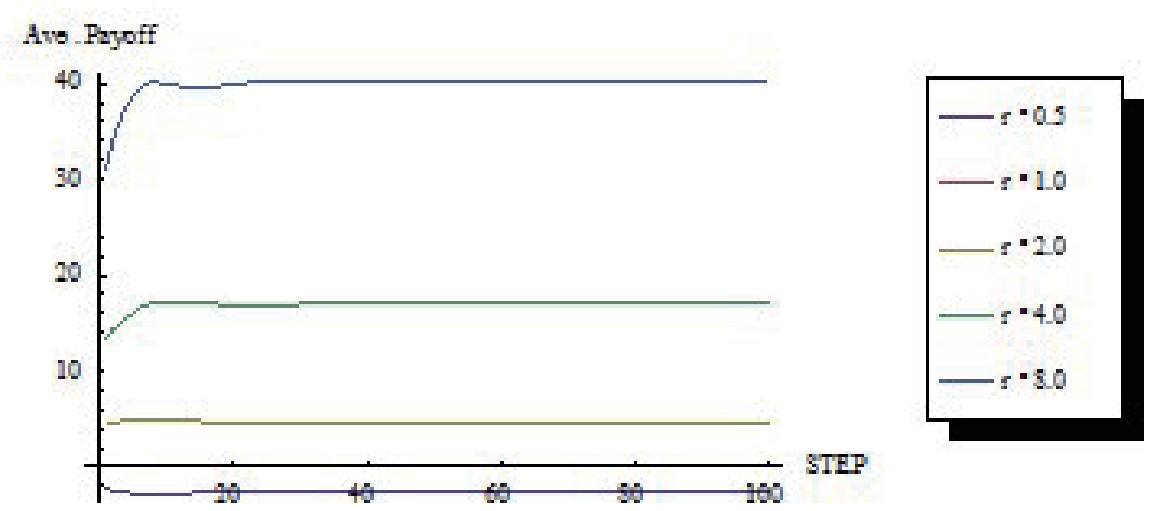}
  \end{center}
  \caption{\small The D ratio in A2-model (SFnet $k=6$) with topology change}
  \label{fig:one}
 \end{minipage}
  \hspace*{3mm}
 \begin{minipage}{0.5\hsize}
  \begin{center}
\includegraphics[width = 7.0cm]{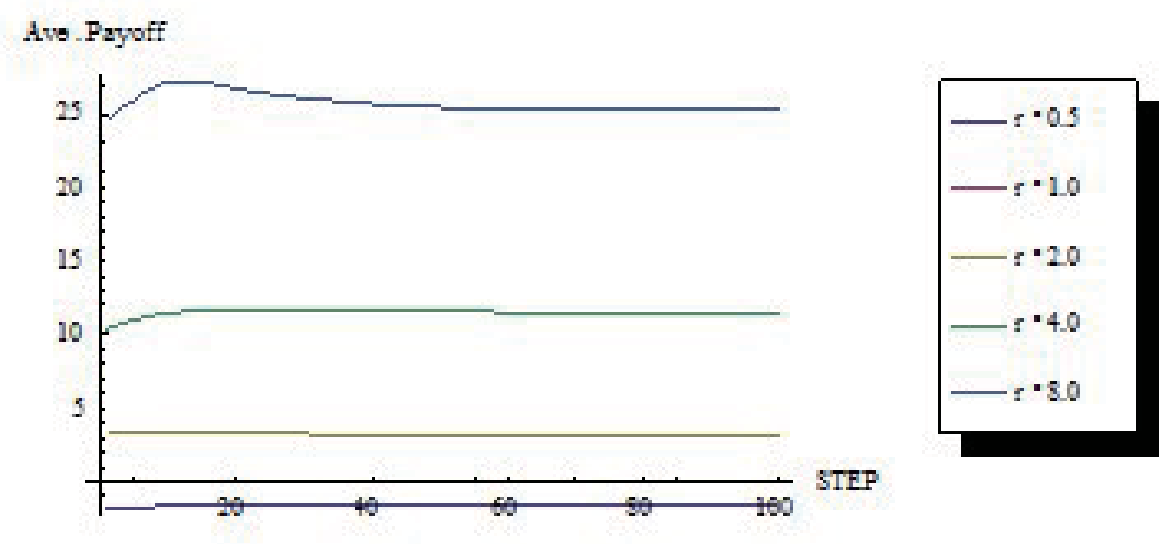}
  \end{center}
  \caption{\small The D ratio in A2-model (SFnet $k=6$) without topology change}
  \label{fig:two}
 \end{minipage}
\end{figure}

 \begin{figure}[bp]
 \begin{minipage}{0.5\hsize}
  \begin{center}
\includegraphics[width = 7.0cm]{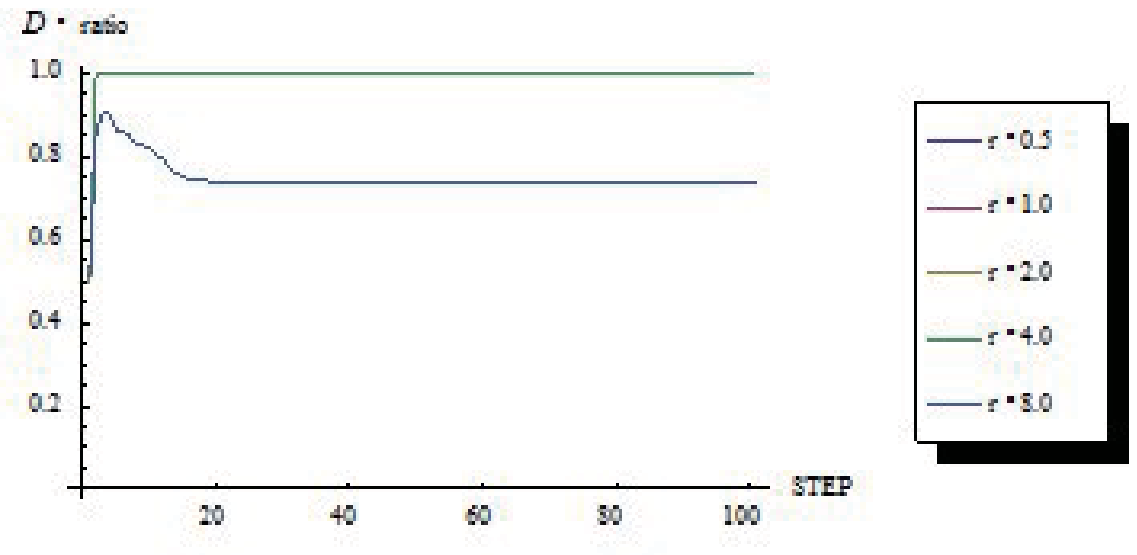}
  \end{center}
  \caption{\small The D ratio in B1-model (ERnet $k=16$) with topology change}
  \label{fig:one}
 \end{minipage}
  \hspace*{3mm}
 \begin{minipage}{0.5\hsize}
  \begin{center}
\includegraphics[width = 7.0cm]{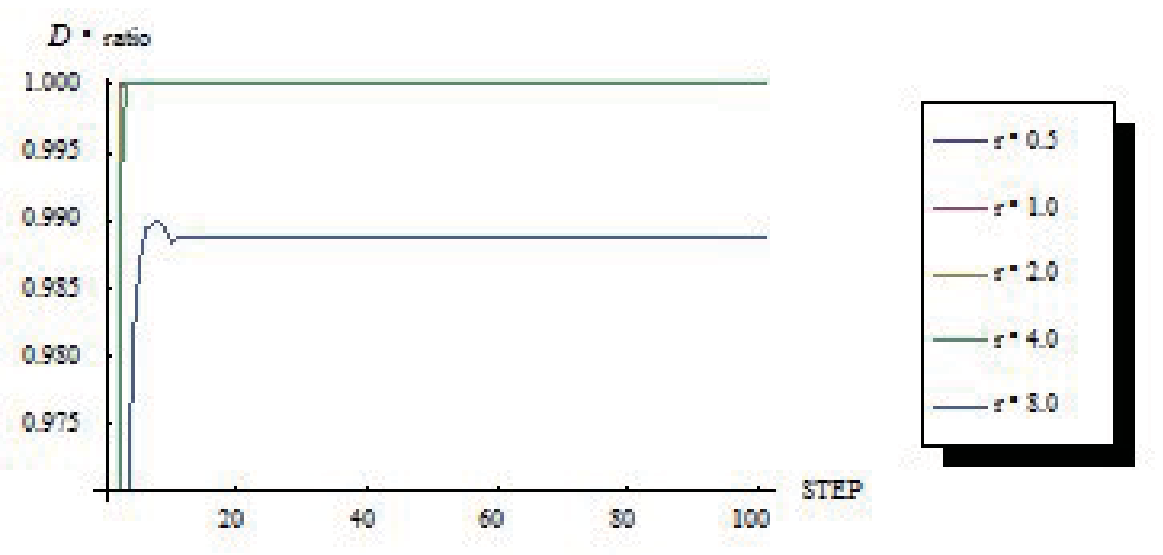}
  \end{center}
  \caption{\small The D ratio in B1-model (ERnet $k=16$) without topology change}
  \label{fig:two}
 \end{minipage}
\end{figure}

 \begin{figure}[bbp]
 \begin{minipage}{0.5\hsize}
  \begin{center}
\includegraphics[width = 7.0cm]{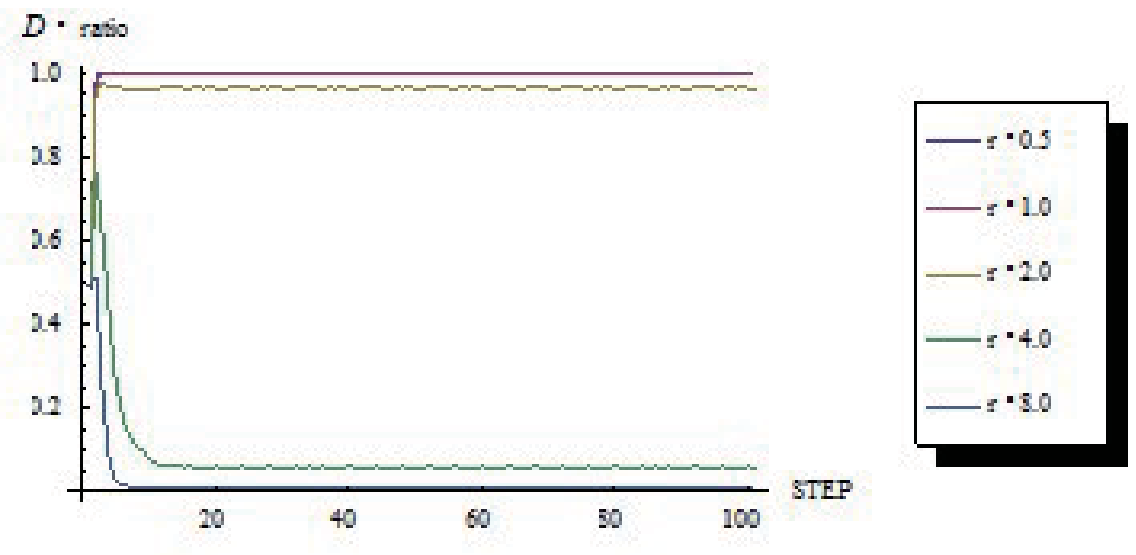}
  \end{center}
  \caption{\small The D ratio in B1-model (WSnet, $w=0.1$ $k=4$) with topology change}
  \label{fig:one}
 \end{minipage}
  \hspace*{3mm}
 \begin{minipage}{0.5\hsize}
  \begin{center}
\includegraphics[width = 7.0cm]{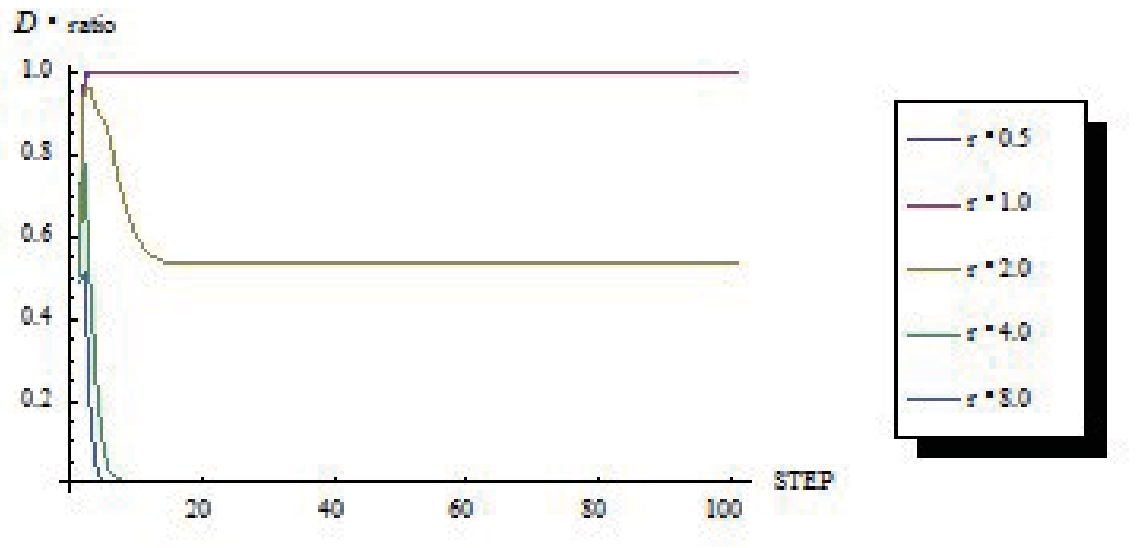}
  \end{center}
  \caption{\small The D ratio in B1-model (WSnet, $w=0.1$ $k=4$) without topology change}
  \label{fig:two}
 \end{minipage}
\end{figure}

 \begin{figure}[bp]
 \begin{minipage}{0.5\hsize}
  \begin{center}
\includegraphics[width = 7.0cm]{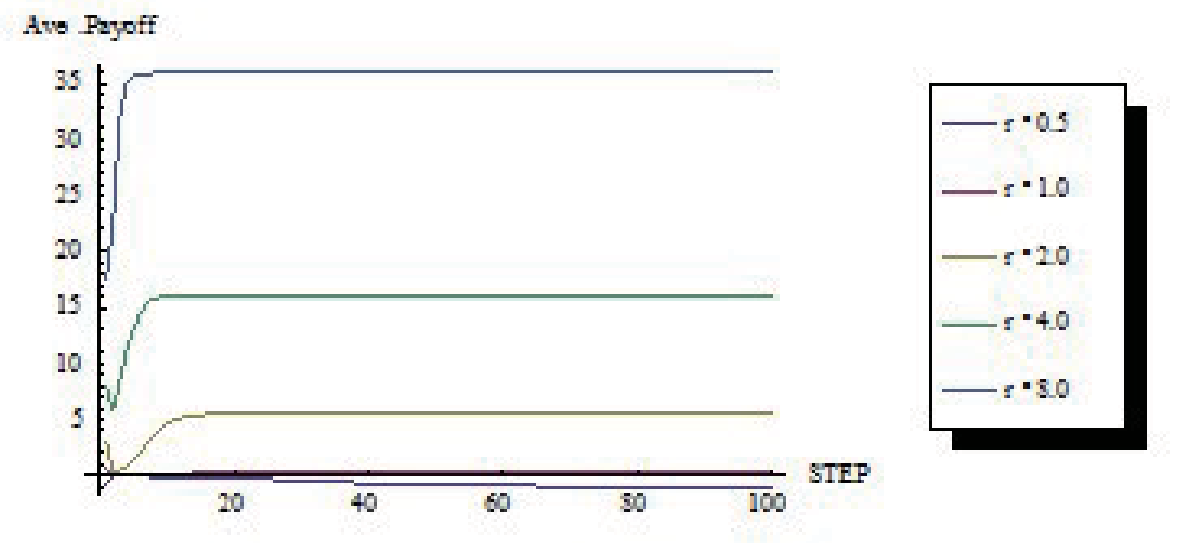}
  \end{center}
  \caption{\small The average payoff in B1-model (ERnet, $k=4$) with topology change}
  \label{fig:one}
 \end{minipage}
  \hspace*{3mm}
 \begin{minipage}{0.5\hsize}
  \begin{center}
\includegraphics[width = 7.0cm]{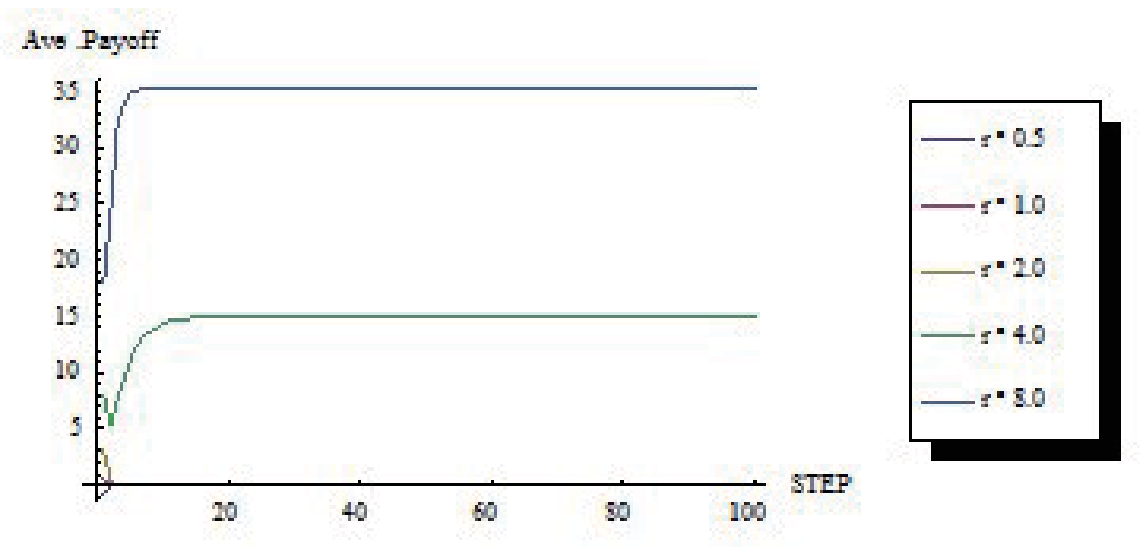}
  \end{center}
   \caption{\small The average payoff in B1-model (ERnet, $k=4$) without topology change}
  \label{fig:two}
 \end{minipage}
\end{figure}

\textbf{B-2model:}Roughly D-strategy is dominant in many cases. 
It, however, is difficult to show the definite and typical results, since the results of simulations strongly depend on the initial network and the initial distribution of players with  C-strategy and D-strategy  so that they are rather unstable.  

\textbf{B-3model:} 
C-strategy is more promoted only in the cases which topology change does not occur.  
Though D-strategy has a majority in the model with topology change  for all $r$ as seen in Fig.16, 
C-strategy gets a majority for some $r$ in the cases without topology change in Fig.17. 
This behavior is contrary to A-models. 
We can not, however, observe the essential differences in both models for the average payoffs as seen in Fig.18 and Fig.19.

 \begin{figure}[tbp]
 \begin{minipage}{0.5\hsize}
  \begin{center}
\includegraphics[width = 6.0cm]{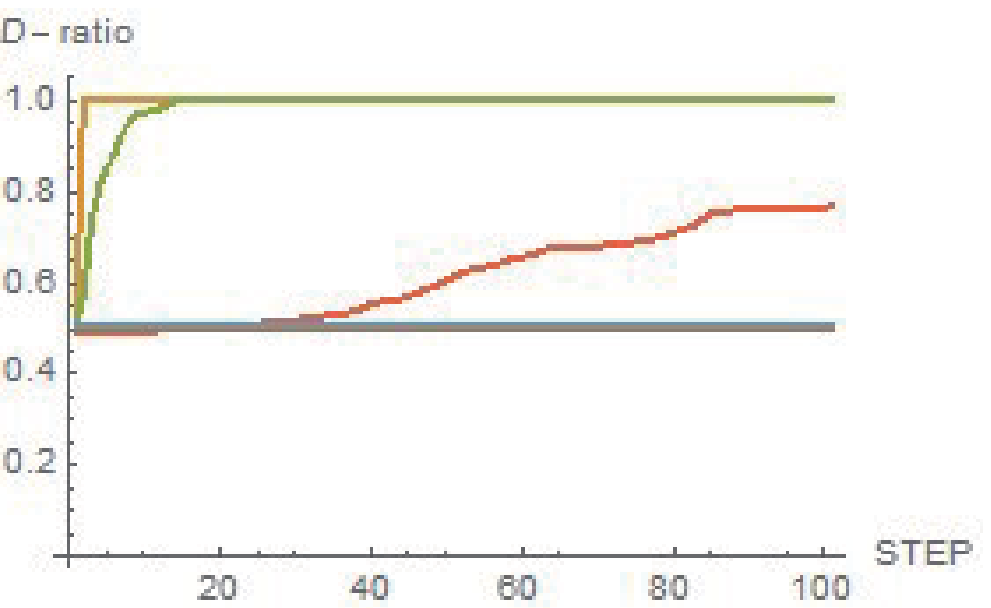}
  \end{center}
  \caption{\small The D-ratio in B3-model (ERnet, $k=4$) with topology change}
  \label{fig:one}
 \end{minipage}
  \hspace*{3mm}
 \begin{minipage}{0.5\hsize}
  \begin{center}
\includegraphics[width = 6.0cm]{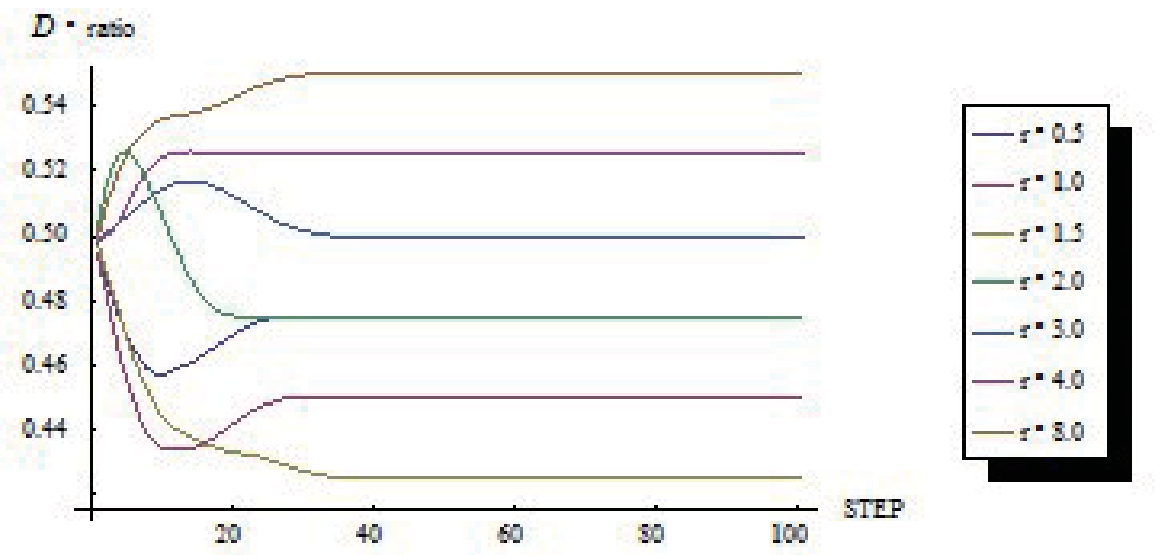}
  \end{center}
  \caption{\small The D-ratio in B3-model (ERnet, $k=16$) without topology change}
  \label{fig:two}
 \end{minipage}
\end{figure}
 \begin{figure}[tbp]
 \begin{minipage}{0.5\hsize}
  \begin{center}
\includegraphics[width = 6.0cm]{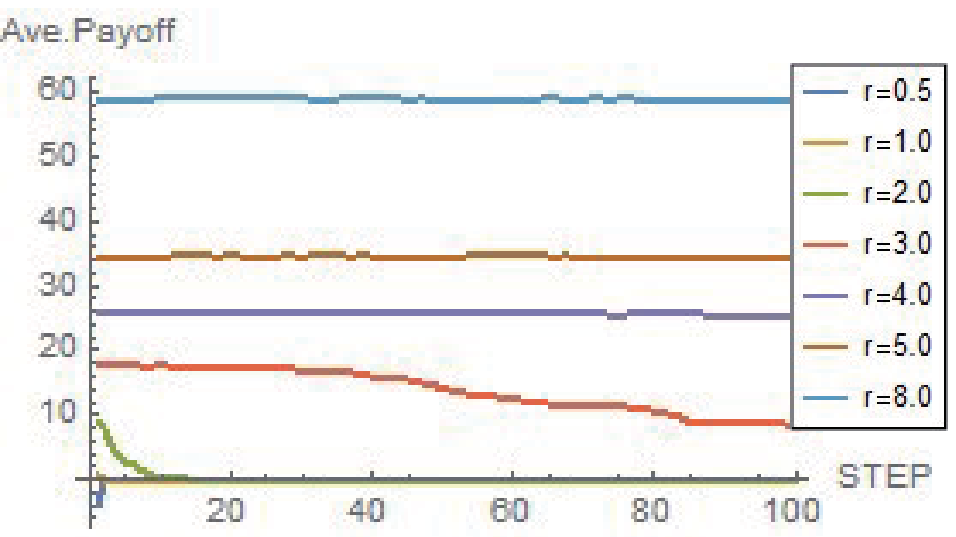}
  \end{center}
  \caption{\small The average payoff in B3-model (ERnet, $k=16$) with topology change}
  \label{fig:one}
 \end{minipage}
  \hspace*{3mm}
 \begin{minipage}{0.5\hsize}
  \begin{center}
\includegraphics[width = 6.5cm]{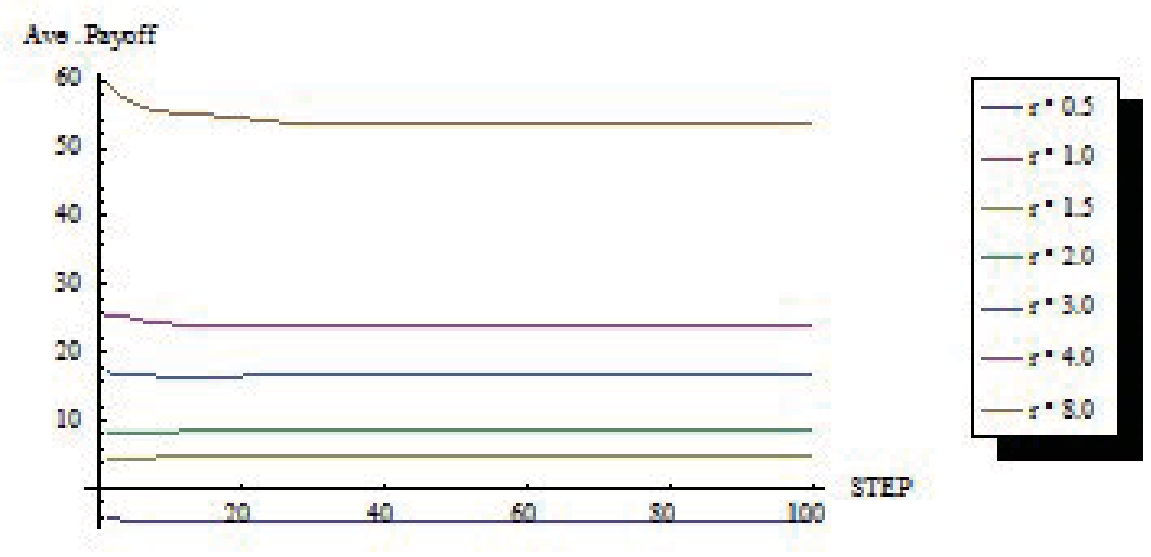}
  \end{center}
  \caption{\small The average payoff in B3-model (ERnet, $k=16$) without topology change}
  \label{fig:two}
 \end{minipage}
\end{figure}

\subsection{Degree Distribution}
In $\alpha$ models, degree distributions come close to the Poisson distribution as random networks, regardless of $r$, for all initial networks. 
An example of these facts is shown in Fig. 20 where the point graph shows the Poisson distribution.   
The nodes with zero degree, however, increase in cases with the low average degree as shown in Fig.21,  
where nodes with D-strategy are almost isolated.

 \begin{figure}[tbp]
  \begin{minipage}{0.5\hsize}
  \begin{center}
\includegraphics[width = 6.0cm]{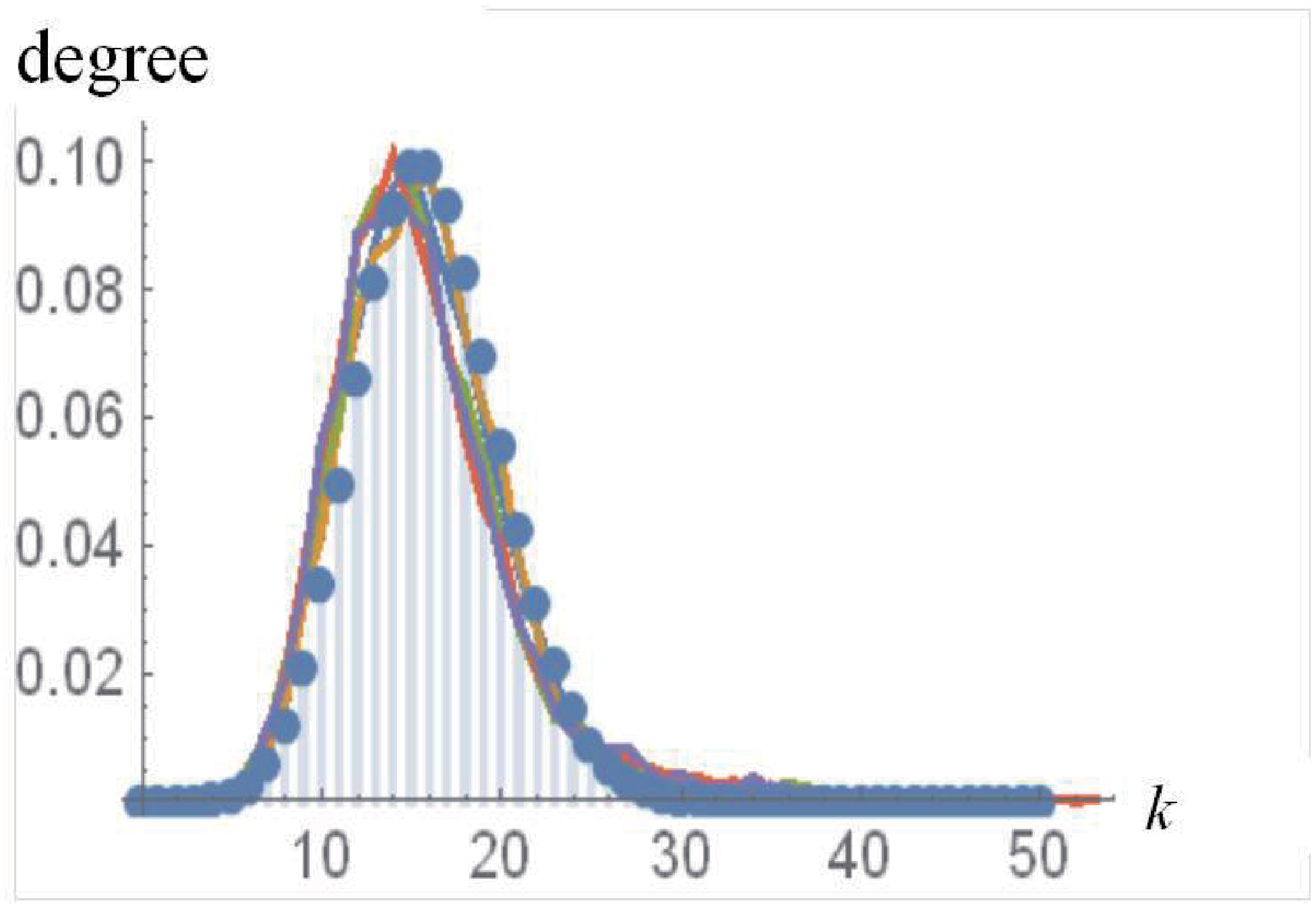}
  \end{center}
  \caption{\small The degree distribution in B1-model (SFnet, $k=4$ ) with topology change}
  \label{fig:one}
 \end{minipage}
  \hspace*{3mm}
 \begin{minipage}{0.5\hsize}
  \begin{center}
\includegraphics[width = 6.0cm]{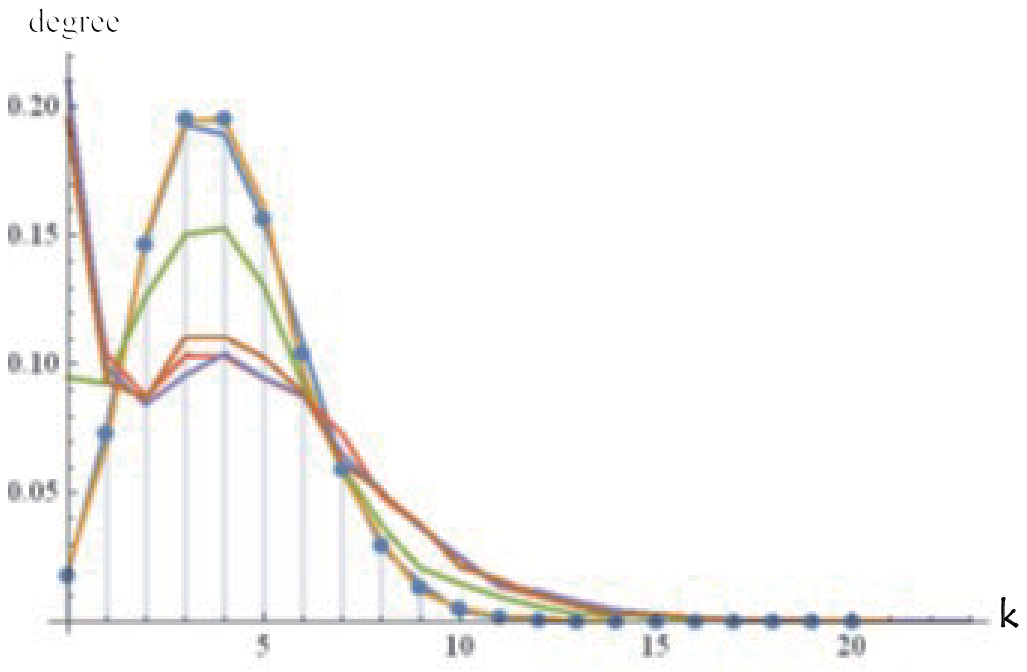}
  \end{center}
  \caption{\small The degree distribution in B1-model (ERnet, $k=4$ )  without topology change}
  \label{fig:two}
 \end{minipage}
\end{figure}
Since the average payoff and D-ratio in $\beta$-models were almost the same as in $\alpha$-models, we omit the details of them. 
But there are apparent differences in the degree distributions in the both models. 
These are shown in Fig.22 and Fig.23. 
The degree distributions at small $k$ basically become line graphs shown in Fig.22 in A-models and also B1-model 
where players with D-strategy who account for a large portion are almost  isolated and have zero degree. 
On the other hand, at large $k$ many players with D-strategy are often isolated but the degree distribution has a small peak at the  original average degree in B2-model and also B3-model. 
Such the situations are not the characteristic of WS-net.  
These characteristics do not depend on initial networks. 
In $\beta$-model, when a player does not have the neighbor of his/her neighbor, the total number of edges decreases. 
Especilally it is easy to fall into such situation for players with D-strategy, who are often broken with other players.

The structures of final networks after network evolution depend heavily on how to attach new edges to nodes but not to cut edges from these facts.

\subsection{Effects of dynamics and Topology change}
The models subjected to the influence of game dynamics are A1, B1 and B3 models. 
On the other hand A2 and B2 models are not subbjected to the influence of such game dynamics. 
Though both  A1 model and A-2 model maintain C-strategy by a certain ratio in the both cases with topology change ($p=q=1$)  and without topology change($p=q=0$),  D-strategy is dominant in the cases without topology change. 
C-strategy is notably dominant in A2-model and  A1-model at large $r$ with topology change.     
The average payoffs of cases with topology change is larger than ones without topology shange. 
It is considered that there are detectable effects of topology change on the A-models with strategy vectors in the average payoff and D-ratio from such observations. 

We obseve different effects as for topology change in B3-model from other models. 
Though D-strategy is dominant on the whole in the cases with topology change, C-strategy mostly gains a majority   in the many cases at large $r$ without topology change. 
In B1-model, C-strategy is dominant at large $r$ and small $k$ but D-strategy is dominant at small $r$. 
In such a way, a kind of phase transition like phenomenon emerges by changing $r$.  
There is, however, not the marked influence of topology change on the average payoff. 
As result, there is not the great influence of topology change in B1-model. 
In B2-model that does not reflect game dynamics, D-strategy is almost dominant regardless of whether there has been topology change or not.  

Accordingly, the topology change has a great infulence on D-ratio and the average payoff in A-models using strategy vectors.   
In B-models, however, it has only a slight influence on the average payoff.  
The influence on the ratio of C-strategy is model dependent. 

Next we consider the influences of the game dynamics. 
There are no crucial defferences between A1-model and A2-model in  D-ratio and the average payoff,  
 regardless of whether there are topology change or not.  
 The qualitative differences between B2-model and B3-model are not observed in the cases with topology cahnge. 
A kind of the phase transition like phenomenon is observed in B1-model at small $k$  by changing $r$. 
The cases without topology change are as follows.  
C-strategy is dominant in many cases of B3-model for large $r$ but D-strategy is mostly dominant in B2-model, especially at small $k$.  
The tendency that the game dynamics promotes C-strategy in the case without topology change is strong. 
Thus there are some  differences in results between the models reflecting the game dynamics and not reflecting it. 
  Based on the above observations, in many cases there are no crucial differences between the models reflecting the game dynamics and not reflection it, but there is only a big difference among B-models that employ the fixed strategy in the cases without topology change.  

 \begin{figure}[tbp]
 \begin{minipage}{0.5\hsize}
  \begin{center}
\includegraphics[width = 6.0cm]{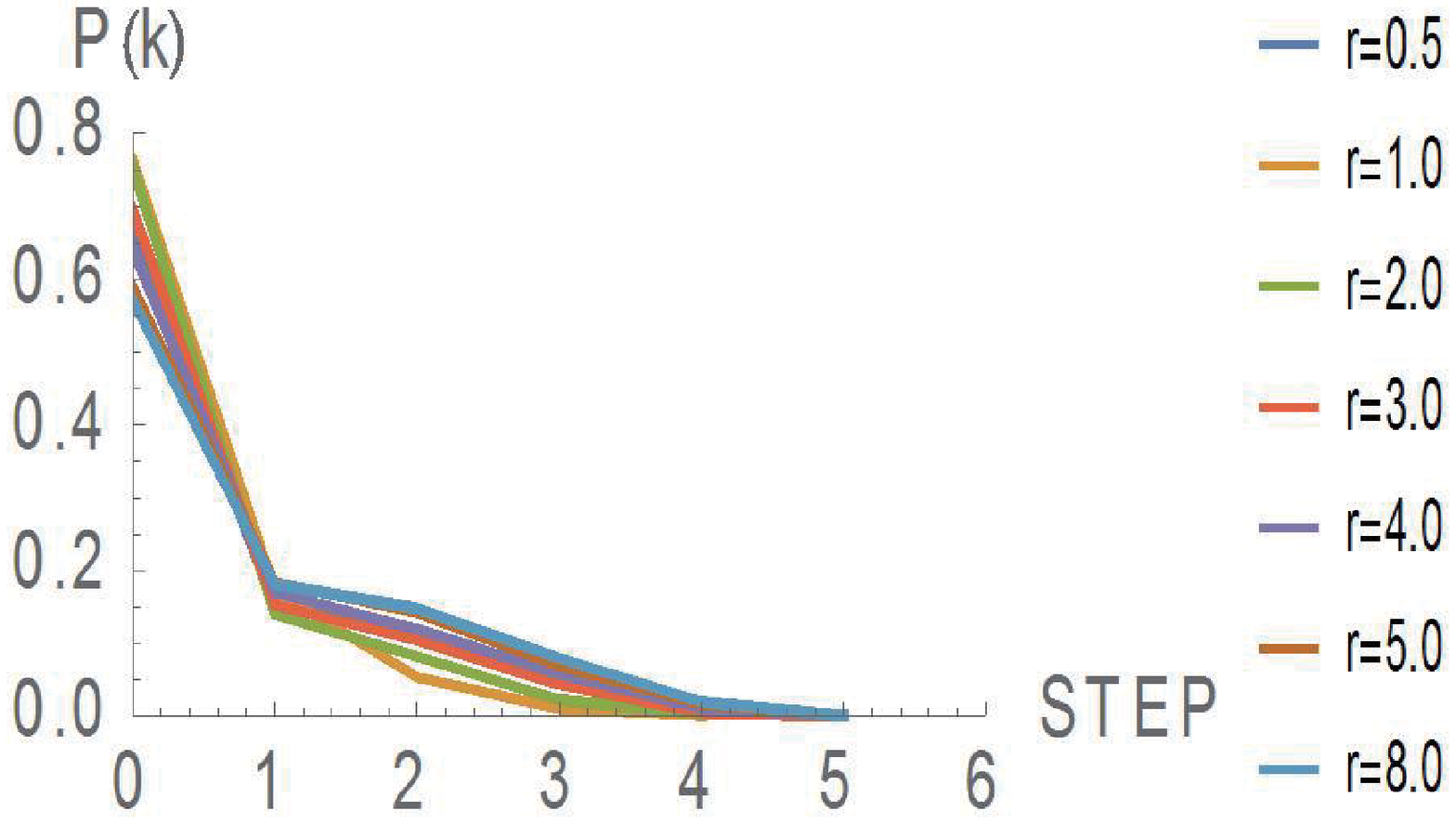}
  \end{center}
  \caption{\small The degree distribution in A1-model (WSnet, $k=4$ and $w=0.01$ with topology change}
  \label{fig:one}
 \end{minipage}
  \hspace*{3mm}
 \begin{minipage}{0.5\hsize}
  \begin{center}
\includegraphics[width = 6.0cm]{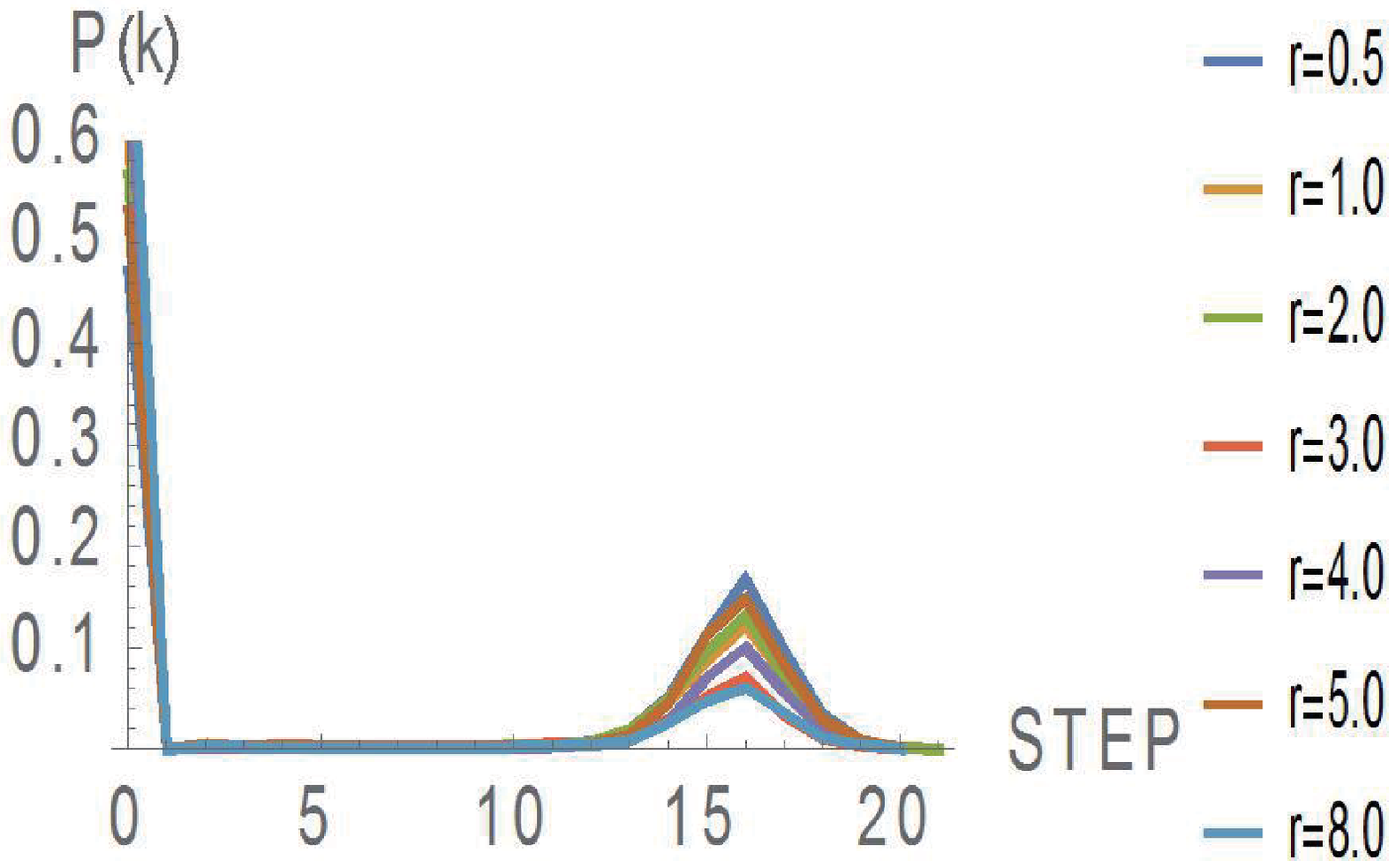}
  \end{center}
 \caption{\small The degree distribution in B2-model (WSnet, $k=16$ and $w=0.1$ with topology change}
  \label{fig:two}
 \end{minipage}
\end{figure}

\subsection{Summaries of Simulations}
On the whole the differences of the initial networks have not influenced on the essential results. 
The final network structures are almost the same especially for all $\alpha$-models, and the degree distribution is mostly Poisson distribution, because it is thought that the way that cut and paste edges for every model is common. 
The average degree of an initial network, however, influences on the degree distribution.  
Thus we think that the  curcial difference by the game dynamics in the final networks has not arisen.  

 There are hardly any defferences in the ratio of the players with C-strategy and the average payoff between $\alpha$-model and $\beta$-model.  
A big difference between the both models arises in the final network structure. 
This shows that a final network structure mainly depends on the way to paste edges between nodes or players.    

When the average degree of a network is small, that is, there are only few participants of the game,  C-starategy is more promoted. 
This shows that the effect that cuts edges to unprofitable players is larger at small $k$.   
Since the players whose edges are cut  is easily isolated at small $k$, they can not frequently get any payoff from PGG. 
The fact that the isolated players increase at small $k$ is supported also from degree distributions. 

The models that preserve or promote C-strategy are  B3 model without topology change and B1-model with small $k$ and large $r$ among B-models with fixed strategies in $\alpha$ model. 
 On the other hand, A1-model and A-2 model and B1-model with  large $r$ and small $k$  preserve or promote C-strategy among the models with topology change. 
Then A2-model using strategy vectors does expressly promotes C-strategy. 
Thus the models using strategy vector tend rather to promote C-strategy, when the topology of networks changes.

\section{Summary}
In this research, we studied the ratio of players with D-strategy,   the average payoff of all players and the final degree distributions by simulating  evolutionary PGG, where players have various types of tactics, on  complex networks. 
The diverse tactics are considered to model the actions that usual persons would take.  
We investigated whether the interacton between the game dynamics and topology change, or a combined effect thereof promotes cooperation and  make people (players) wealthy or not.  
 As a result, it is shown that D-strategy is dominant in many cases, dependently on $r$. 
 C-strategy, however, rather tends to comparatively flourish when $k$ is small than when large $k$.  
It, however,  is almost impossible that the interaction or the combined effect of the network evolution and PGG dynamics promote cooperation. 
A few exceptions that the combined effect somewhat preseves C-strategy are only A1-model and B1-model with small $k$ and large $r$.  
Especially A2-model remarkably promote C-strategy but it is independent of the game dynamics. 
Thus the  increment of C-strategy is not due to the interaction or the combined effect but only due to topological evolution.   
In contrast,  
D-strategy is obviously dominant in the models without topology change and PGG dynamics. 
Other factor such as punishment for promoting so much C-strategy may be needed\cite{And},\cite{Perc8}.  

It is shown that the degree distribution of initial networks except random networks varies considerably to approach Poisson-like distribution in $\alpha$-model. 
We found that the degree distribution of finally constructed networks depends chiefly on the new connection rule 
of edges among topology change by comparing  $\alpha$-model  and $\beta$-model. 

Diverse people coexist in the real world. 
We should make more realistic models, where players can employ diverse tactics simultaneously,A1, A2, B1 $\cdots$ and so on, to reflect such the situation. 
 The results in this article are preliminary ones, where some models with tactics probable in the world are considered in a piecewise manner. 
Deeper considerations with more general modeling are therefore indicated in subsequent studies.

\end{document}